\documentclass[12pt,twoside]{article}
\usepackage{fleqn,espcrc1}

\usepackage{epsfig}

\newcommand{\be}{\begin{equation}}
\newcommand{\ee}{\end{equation}}
\newcommand{\ba}{\begin{array}}
\newcommand{\ea}{\end{array}}
\newcommand{\baa}{\begin{eqnarray}}
\newcommand{\btab}{\begin{tabular}}
\newcommand{\etab}{\end{tabular}}
\newcommand{\eaa}{\end{eqnarray}}

\newcommand{\ci}[1]{\cite{#1}}

\newcommand{\lab}[1]{\label{#1}}
\newcommand{\re}[1]{\ref{#1}}

\newcommand{\PRD}[3]{Phys.\ Rev.\ D {\bf {#1}}, {#2} ({#3})}

\newcommand{\NPB}[3]{Nucl.\ Phys.\ B {\bf {#1}}, {#2} ({#3})}
\newcommand{\PLB}[3]{Phys.\ Lett.\ B {\bf {#1}}, {#2} ({#3})}

\newcommand{\EJP}[3]{Eur. J. Phys.\ C {\bf {#1}}, {#2} ({#3})}

\begin{document}

\begin{titlepage}
\begin{flushright}
\begin{tabular}{l}
TUM/T39-99-16\\
\end{tabular}
\end{flushright}
\vskip0.4cm

{\large\bf Conformal string operators and 
skewed parton distributions}\footnote{Talk given at the XVth International
Conference on Particles And Nuclei, June 10-16, 1999 Uppsala, Sweden}   
\\[4mm]
\begin{center}
     N. Kivel\footnote{Alexander von Humboldt fellow, on leave from
Petersburg Nuclear Physics Instite, Petersburg, Russia.}   \\
                            
%
{\small \em 
Physik Department, Technische Universit\"{a}t M\"{u}nchen, 
Germany. }
\end{center}

\vspace{3cm}

\centerline{\bf Abstract}
\begin{center}
\begin{minipage}{15cm}

We discuss skewed parton distributions in the coordinate space. 
Solution of the corresponding LO evolution equation is constructed in terms of
eigenfunctions of the evolution kernel 
and its relation to the conformal symmetry is explained.

\end{minipage}
\end{center}

\end{titlepage}

Recently, there has been a renewed interest in QCD evolution of skewed parton
distributions \cite{Ji97Rad97}. Skewed parton distributions play a
crucial role in description of hard, exclusive QCD processes which are 
actively considered  as tools for investigation of new aspects of
non-perturbative QCD dynamics. However, it is clear that
before  non-perturbative information can be reliably extracted from
experimental data, all perturbative aspects, such as QCD evolution, have to be
understood.  So far, the main effort has been devoted to studies of evolution 
of the skewed parton distributions (SPD) in the momentum representation 
\cite{FFGS97}.

SPD are defined through matrix elemets of  twist-2 
string operators. Consider as an example a nonsinglet quark operator:
\be
\lab{eq:Q_op_def}
O(\alpha,\beta) = {\bar q}({\textstyle \frac{\alpha+\beta}{2}}z) \hat{z}
{\cal P} \exp \left\{
 -i g \-\int_{\frac{\alpha+\beta}{2}}^{\frac{\alpha-\beta}{2}} 
z_\mu A^{\mu}(tz) dt \right\}
q({\textstyle\frac{\alpha-\beta}{2}}z)\, ,
\ee
Corresponding SPD can be introduced in the following way
\be 
\left \langle P'\right| 
O(\alpha,\beta) 
\left|P\right \rangle
= 
\bar N(P')\,\hat z\, N(P) e^{-i \alpha \frac{r}{2}\cdot z}
\!\int_{-1}^1 \!\! du \,F(u,\xi;\mu^2) e^{i u\beta({\bar P}\cdot z)}
 + ...
\label{def:corr_F_xi}
\ee
where dots denote other Dirac structures. 
 $N(P)$ and $\bar N(P')$ denote initial and final nucleon spinors,
respectively. The average nucleon momentum is denoted by $\bar P = (P +
P')/2$, and the momentum transfer is $r = P-P'$. The asymmetry parameter $\xi$
is defined by $r\cdot z = 2 \xi {\bar P}\cdot z$.

Scale dependence of the SPD is governed by a generalised evolution equation
\be
\mu \frac{d}{d\mu}F(x,\xi;\mu^2)=\frac{\alpha_S}{4 \pi}
\int_{-1}^{1} dy V(x,y,\xi)F(y,\xi;\mu^2)
\lab{evol_eq}
\ee
At the LO, the evolution kernel $V(x,y,\xi)$  has a set of eigenfuctions
associated with local conformal operators:
\be
\int_{-1}^{1}C^{3/2}_j(x/\xi) V(x,y,\xi)dx =\gamma_j C^{3/2}_j(y/\xi)
\lab{eigenC}
\ee
As these eigenfuctions do not form a complete set outside the region
$|x/\xi |>1$, they can not be used for expansion of the SPD.
This can also be  understood  by the hybrid properties of the SPD.
Let us split $F(x,\xi,\mu^2)$ in two pieces:
\be
F(x,\xi,\mu^2)=F_<(x,\xi,\mu^2)+ F_>(x,\xi,\mu^2)
\ee
with
\be
F_<(x,\xi,\mu^2)=\theta(x<\xi)F(x,\xi,\mu^2),
\mskip10mu 
F_>(x,\xi,\mu^2)=\theta(x>\xi)F(x,\xi,\mu^2)
\lab{sg}
\ee
$F_<,\ F_>$ describe partons with $x<\xi$ and $x>\xi$, respectively. The
crucial point is that the evolution of $F_<$ and $F_>$ is qualitatively
different \ci{Ji97Rad97,KivMan99}.  Partons which at the initial scale belonged
to the segment $0 \le u \le \xi$ stay there in the course of the evolution. On
the other hand, partons which belonged initially to the segment $\xi < u \le 1$
diffuse into the segment $0 \le u \le \xi$ and never come back.  Mathematically
this means that the function $F_<(x,\xi,Q^2)$ will be restricted to the initial
region $0 \le u \le \xi$ but $F_>(x,\xi,Q^2)$ will expand to whole interwal $0
\le u \le \xi$.  The former and latter cases resemble the ERBL and DGLAP
evolution, respectively.  As it follows, properties of SPD in the region
$x>\xi$ SPD are similar to forward parton distribution $f(x)$, while in the
region $x<\xi$ SPD looks like a distribution amplitude. Expansion in the
ortognal set eigenfunctions (\ref{eigenC}) is valid for the ERBL-region $x<\xi$
only, and reflects a typical structure of evolution for such configurations of
partons. As we see, in momentum space there are two different regions in $x$
with different evolution properties.

The situation is different in coordinate space. Coordinate-space 
SPD is defined through a Fourier transformation:
\be
{\cal F}(\beta,\xi;\mu^2) = \frac{1}{\pi}\!\int_{-1}^1 \!\! dx
\,F(x,\xi;\mu^2) \, e^{i x\beta}
\lab{Four}
\ee
It is easy to see that, unlike in the momentum space, the coordinate-space SPD
${\cal F}_{<,>}$ associated with functions (\ref{sg}) are defined in the
same interval $0\le \beta \le \infty$.  So, one can hope that an ortogonal set
of coordinate-space eigenfunctions exists and  can be used as a basis for
expansion of SPD.

Evolution equations in coordinate space have been discussed e.g., 
in \ci{BalBra88}, where the role of the classical conformal symmetry
was emphasized. The authors of \ci{BalBra88} were able
to write the solution in form of complex integral over conformal
spin $j$. Recently, in \ci{KivMan199} the solution have been obtained
in terms of coordinate-space eigenfuctions corresponding to integer $j$.   
Here we will obtain solution for the SPD in coordinate space using a 
formal trick and then explain its relation with conformal symmetry.

Recall that a function $f(x)$ can be
expanded in a Neumann series according to \cite{Erdeylyi}
\be\lab{eq:Neumann_series}
f(x) = \sum_{n=0}^\infty (2 \nu + 2 n ) J_{\nu + n}(x) \int_0^\infty
\frac{d\lambda}{\lambda}\, f(\lambda) J_{\nu + n}(\lambda) \, .
\ee
In particular, one finds that $ e^{i x\beta}$ 
can be decomposed according to the Sonine's
formula \cite{Erdeylyi}:
\be\lab{eq:Neumann_exp}
 e^{i x \beta} = 
\left(\frac{2}{\beta \xi}\right)^\frac{3}{2} \, 
\Gamma\left[3/2\right] \,\sum_{n=0}^\infty\, i^n (3/2 + n)\, 
C^{\frac{3}{2}}_n({x}/{\xi})\, 
J_{\frac{3}{2}+n}(\beta \xi) \,
\ee
Inserting this expansion in the definition of the coordinate-space skewed quark
distribution (\re{Four}) and interchanging summation and
integration 
one obtains:
\be\lab{eq:CSD_s_exp_bess}
{\cal F}(\beta,\xi;\mu^2) =
\frac{1}{\sqrt{\pi}}\,\left(\frac{2}{\beta \xi}\right)^\frac{3}{2}\,
\sum_{n=0}^\infty i^n ({3}/{2}+n) J_{\frac{3}{2}+n} (\beta \xi)
\int_0^1 \!\! du
\,F(\omega,\xi;\mu^2) \, C^\frac{3}{2}_{n}({\omega}/{\xi}) \, .
\ee
Now, note that a Gegenbauer moment
is proportional
to the matrix element of multiplicatively renormalizable local conformal
operator and its scale dependence is therefore given by  
$$
\int_0^1 d\omega F(\omega,\xi;Q^2)C^\frac{3}{2}_{n}({\omega}/{\xi}) = 
L_{1+n}\int_0^1d\omega F(\omega,\xi;\mu^2)C^\frac{3}{2}_{n}
({\omega}/{\xi}),\ \  
L_{k} =
\left(\frac{\alpha_S(\mu)}{\alpha_S(Q)}\right)
^{-\frac{\gamma(k)}{b_0}} 
$$
%
As it follows, the scale dependence of the coordinate space distribution  
${\cal F}(\beta,\xi;\mu^2)$ is given simply by
\be\lab{eq:CSD_s_scale_dep}
{\cal F}(\beta,\xi;Q^2) =\!\!
\frac{1}{\sqrt{\pi}}\,\left(\frac{2}{\beta \xi}\right)^\frac{3}{2}\,
\sum_{n=0}^\infty i^n ({\textstyle \frac{3}{2}}+n) 
L_{n + 1}
J_{\frac{3}{2}+n} (\beta \xi)
\int_0^1 \!\! d\omega
\,F(\omega,\xi;\mu^2) \, C^\frac{3}{2}_{n}({\omega}/{\xi}) \, .
\ee
Now we show that equations (\ref{eq:CSD_s_scale_dep}), 
can be naturally understood as expansions of
coordinate-space skewed quark distributions in terms of matrix elements of
non-local, multiplicatively renormalizable, conformal operators. Indeed,
applying (\re{eq:Neumann_series}) one can rewrite (\ref{eq:CSD_s_scale_dep})
 as a Neumann-type series:
\be
\lab{eq:CSD_scale_dep_nonl}
{\cal F}(\beta,\xi;Q^2) =\!\! \beta^{-\frac{3}{2}}
\sum_{n=0}^\infty (3+2n) 
J_{\frac{3}{2}+n} (\beta \xi)
L_{ n + 1}
\int_0^\infty d\lambda \sqrt{\lambda}
\,{\cal F}(\lambda,\xi;\mu^2) \, J_{\frac{3}{2}+n}(\lambda \xi)
\ee
We are now in the position to make the relation to conformal symmetry explicit.
Let us start from the obvious identity
\be\lab{eq:identity}
O(\alpha,\beta) = \int_{-\infty}^\infty d\alpha^\prime  \int_0^\infty
d\beta^\prime\, \delta(\alpha-\alpha^\prime) \delta(\beta-\beta^\prime)
O(\alpha^\prime,\beta^\prime) \, . 
\ee
Applying (\re{eq:Neumann_series}) one finds a representation of a
$\delta$-function in terms of a Neumann series
\be
\beta \delta(\beta-\beta^\prime) = \sum_{n=1}^\infty (1+2n) 
J_{\frac12+n}(\beta)J_{\frac12+n}(\beta^\prime) \, .
\ee
Inserting this expansion into (\re{eq:identity}) 
one finds that the string
operator $O(\alpha,\beta)$ can be decomposed as
\be\lab{eq:conf_exp_1}
O(\alpha,\beta) = \beta^{-\frac{3}{2}}
\int_{-\infty}^\infty \frac{dk}{2\pi}\, e^{-ik\alpha}
\sum_{j=1}^\infty (1+2j) 
J_{\frac{1}{2}+j}(|k|\beta) S({1}/{2}+j,k;\mu^2)
\ee 
in terms of conformal string operators $S(\frac{1}{2}+j,k;\mu^2)$. 
Such operators
\be\lab{eq:conf_op_S}
S({1}/{2}+j,k;\mu^2) = \int_{-\infty}^{\infty} d\alpha \, e^{i k \alpha}\,
\int_0^\infty d\beta \, \sqrt{\beta} J_{\frac{1}{2}+j} (|k|\beta)
\, O(\alpha,\beta) \, ,
\ee
introduced first in
\cite{BalBra88}, form a representation of a conformal group and are therefore
multiplicatively renormalizable at a one-loop level \cite{BalBra88}, i.e.
\be\lab{eq:conf_op_S_reno}
S({1}/{2}+j,k;Q^2) = L_j \, S({1}/{2}+j,k;\mu^2) \, .
\ee
At this point $j$ can be easily identified with the conformal spin.
Taking matrix elements of both sides of the above equations one immediately
reproduces equations (\re{eq:CSD_s_scale_dep}).

Note that in the coordinate-space representation the corresponding LO amplitude
$M(\xi;\mu^2)$ can be written in following way:
\be 
\lab{eq:M_CSD_def}
M(\xi;Q^2) \propto i \pi \!\int_{0}^\infty \!\! d\beta \, e^{-i \beta
\xi} \, {\cal F}(\beta,\xi;Q^2)
\ee  
We have checked, using various models of skewed quark
distributions, 
that the numerical algorithm for evaluation of physical amplitudes, based on
equation (\ref{eq:M_CSD_def}), gives accurate and stable
results, see Figure 1 for an example,
except for a case where the variable $\xi$ becomes small. This is 
related to the observation  that a non-zero $\xi$ provides a natural cut-off 
for large 
$\beta$ behavior of coordinate-space distributions, which significantly
improves the convergence of the Fourier integral (\ref{eq:M_CSD_def}), as
compared to the forward case.
$$
|M(\xi)|^2
\ba{l} 
\epsfig{file=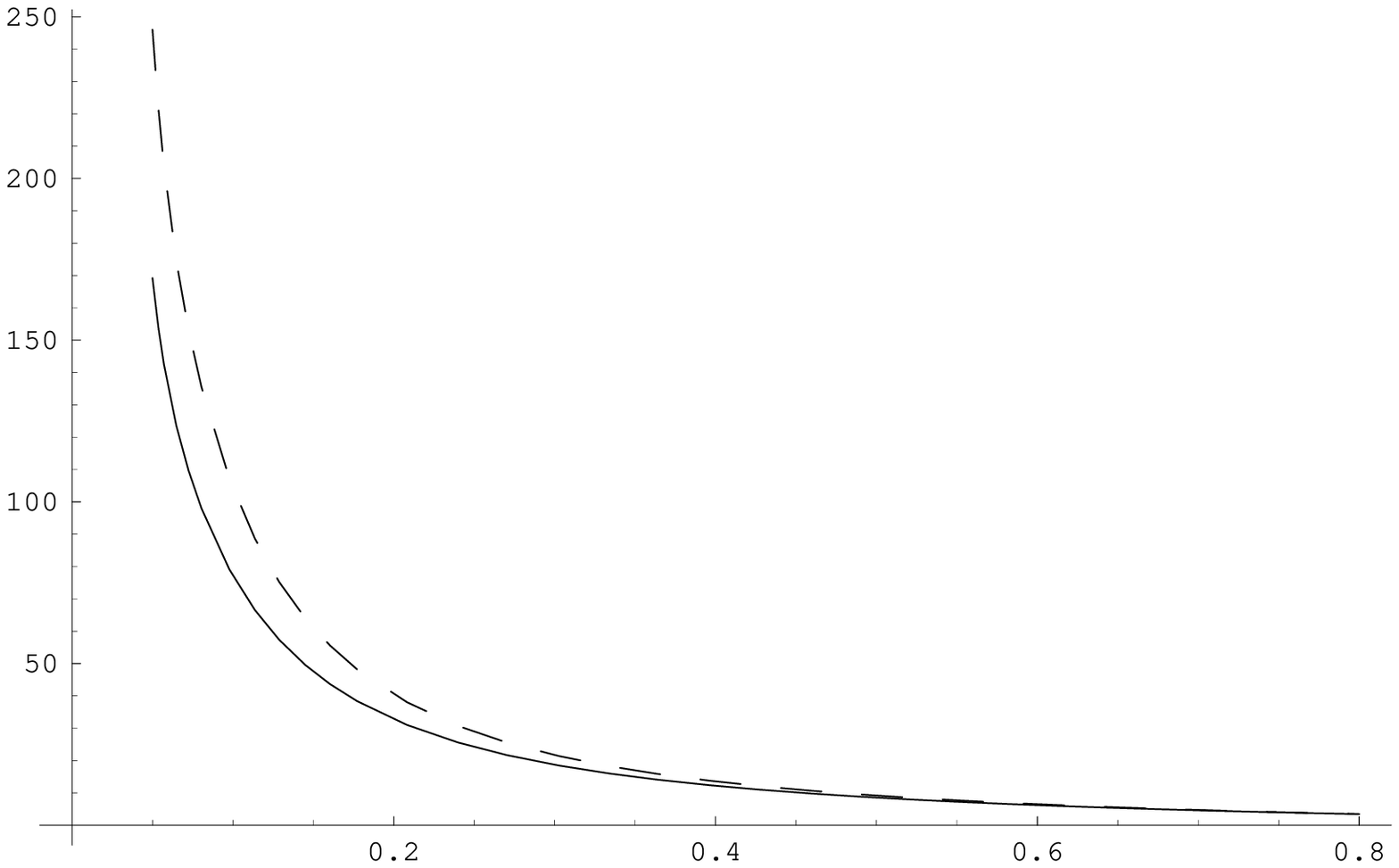,height=60mm,width=66mm}
{}_{{}_{\textstyle\mskip5mu \xi}}
\ea
$$
\vskip1mm
\noindent  
Figure 1. Typical results of evolution of $|M(\xi)|^2$ as a function of
$\xi$, 
starting from a $\xi$-independent initial
conditions $F(u,\xi;\mu_0^2) = 1.1641 u^{-\frac{1}{2}}(1-u)^{3.5}$. The solid
line denotes $|M(\xi)|^2$ at the initial scale  
$\mu_0 = 1.777$ GeV, the dashed line represents $|M(\xi)|^2$ evolved to 
$\mu = 10$ GeV.
\vskip3mm 

\noindent This work has been supported by AvHumboldt Stiftung.

\end{document}